\documentclass[prl,superscriptaddress,showpacs,twocolumn]{revtex4}

\usepackage{graphicx}
\usepackage{graphicx,amsfonts}
\usepackage{epsfig,amsmath}
\usepackage{verbatim}
\usepackage{amssymb}

\def\ba{\begin{eqnarray}}
\def\ea{\end{eqnarray}}
\def\beq{\begin{eqnarray}}
\def\eeq{\end{eqnarray}}
\def\be{\begin{equation}}
\def\ee{\end{equation}}
\def\eg{e.g.}
\def\ie{i.e.}

\def\rmd{{\rm d}}
\def\rme{{\rm e}}
\def\H{{\mathcal H}}
\def\tm{{\tilde M}}

\begin{document}

\title{Dynamic crossover in the global persistence at criticality} 

\author{Raja Paul}
\affiliation{BIOMS, IWR, University of Heidelberg, 69120 Heidelberg, Germany} 
\author{Andrea Gambassi}
\affiliation{Max-Planck-Institut f\"ur Metallforschung, Heisenbergstr. 3,
  70569 Stuttgart, Germany}
\affiliation{Institut f\"ur Theoretische und
Angewandte Physik, Universit\"at Stuttgart, Pfaffenwaldring 57, 70569
Stuttgart, Germany} 
\author{Gr\'egory Schehr}
\affiliation{Laboratoire de Physique Th\'eorique (UMR du
  CNRS 8627), Universit\'e de Paris-Sud, 91405 Orsay Cedex,
  France}


\begin{abstract}
We investigate the global persistence properties of critical systems
relaxing from an initial state with non-vanishing value of the
order parameter (\eg, the magnetization in the Ising model). 
The persistence probability 
of the global order parameter displays two 
consecutive regimes in which it decays algebraically in 
time with two distinct universal exponents. 
The associated crossover is controlled by 
the initial value $m_0$ of the order parameter 
and the typical time at which it occurs diverges 
as $m_0$ vanishes. 
Monte-Carlo simulations of the two-dimensional Ising model with
Glauber dynamics display clearly this crossover. The measured
exponent of the ultimate algebraic decay 
is in rather good agreement with our theoretical predictions for the Ising
universality class. 
\end{abstract}

\maketitle

\section{Introduction}
In spite of many efforts during the last decades, a detailed
description
of the non-equilibrium dynamics of statistical systems is
still lacking. 
In addition to disordered and glassy systems~\cite{leticia_leshouches}
even simpler ones (\eg, pure magnets)
have revealed unexpected non-equilibrium dynamical behaviors such as
aging~\cite{leticia_pure,critical_review}.
Close to {\it critical} points, some aspects of these collective behaviors
become largely independent of the microscopic details of the system 
({\it universality}) and therefore they can be very effectively investigated by
means of simplified models, among which field-theoretical
(FT)~\cite{critical_review} ones.
Within this approach one studies, for instance, the
non-equilibrium relaxation from a
state with non-vanishing value $m_0$ of 
the time-dependent average order parameter $m(t)$ (\eg, the
magnetization of a ferromagnet). It turns out~\cite{janssen_rg} that, after a
non-universal transient, $m(t)$ grows in time as 
$m(t)\propto m_0 t^{\theta'}$ for $t \ll \tau_m \propto
m_0^{-1/\kappa}$ whereas, for $t \gg \tau_m$,  
$m(t)$ decays algebraically to zero as $m(t) 
\propto t^{-\beta/(\nu z)}$. These different time dependences are
characterized by the universal exponents  $\theta'$ (the so-called initial-slip
exponent~\cite{janssen_rg}) and $\kappa = \theta' + \beta/(\nu z)$, where 
$\beta$, $\nu$ and $z$ are the usual static and dynamic (equilibrium) critical
exponents, respectively. 
In addition to the actual time dependence of $m(t)$,
the {\it persistence} properties of thermal fluctuations $\delta m (t)$ 
around $m(t)$ provide useful informations on the dynamics  of the
system (especially on its non-Markovian nature) and indeed in recent
years they have attracted, in various contexts, 
considerable attention both theoretically~\cite{satya_review} and
experimentally~\cite{persistence_exp}. 
For a stochastic process $\{X(t)\}_{t\ge 0}$ with zero mean, 
the persistence probability $P(t)$
is defined as the probability that $X(t)$ does not change sign up to time $t$.
A related quantity which might be used to characterize the process is the
first-crossing (or first-passage) time $t_{\rm cro}$, defined as the
time it takes $X(t)$ to change its sign for the
first time. 
Clearly, the
probability distribution $p(t_{\rm cro})$ is related 
to $P(t)$ via $p(t_{\rm cro}) = - P'(t_{\rm cro})$.
For a ferromagnet at finite temperature $T$
the persistence probability of a single
spin (\ie, of the fluctuations of the {\it local} order parameter) 
decays always exponentially in time due to fast thermal 
fluctuations. The  
long-time decay of the persistence probability
$P_c(t)$ of $\delta m(t)$ 
(\ie, of the fluctuations of the 
{\it global} order parameter), instead, becomes
algebraic upon approaching the critical point: $P_c(t) \propto t^{-\theta}$,
with a non-trivial exponent $\theta>0$. 
This exponent was calculated analytically for some 
ferromagnetic models with different dynamics (Models A and C~\cite{hohenberg77}) and 
random initial
condition~\cite{majumdar_critical,oerding_persist} 
(\ie, $m_0 = 0$)
and for reaction-diffusion systems in the directed
percolation (DP) universality class, initially in the active
phase~\cite{oerding_dirperc} 
(\ie, $m_0=1$ on the lattice, where
the order parameter is the spatial density of reacting particles). 
We shall denote
$\theta_0 \equiv \theta(m_0 = 0)$ and 
$\theta_\infty \equiv \theta(m_0 = 1)$.
In the previous cases analytical progresses rely on the fact that 
the global fluctuating order parameter  $m + \delta m$ has a
Gaussian distribution for all finite times $t$ in the thermodynamic
limit. Indeed, for a $d$-dimensional system of linear size $L$,
$m+\delta m$ is given by the sum of $L^d$ random 
variables (the {\it local} fluctuating degrees of freedom, \eg, spins 
in ferromagnets) which are correlated only across 
a {\it finite}, time-dependent correlation length $\xi(t)$. In
the thermodynamic limit $L/\xi(t) \gg 1$ the 
central limit theorem implies that $m + \delta m$ is a Gaussian
process, for which powerful tools have been developed in order to determine the
persistence exponent~\cite{satya_review,satya_clement_persist}. Remarkably,
under the {\it additional} assumption that the process is Markovian,
$\theta$ can be related 
to known critical exponents, via two 
(hyper)scaling relations
$\theta_0 \equiv \mu_0  
= -\theta' + (1-\eta/2)/z$ (where $\eta$ is the Fisher
exponent) and $\theta_\infty \equiv \mu_\infty =
1+d/(2z)$ which are valid for spin systems and 
for the DP universality class, respectively.
For $m_0 = 0$ (weak) non-Markovian corrections were found
at two-loop and one-loop order in a dimensional expansion around $d=4$, for
Model A and Model C, respectively, in
rather good agreement with Monte Carlo estimates $\theta_0 =
0.237(3)$ in dimension $d=2$ \cite{Schulke97} and $\theta_0 \simeq 0.41$ in
$d=3$~\cite{stauffer96}. 
For completely ordered initial conditions
($m_0 = 1$), instead, these corrections have 
not been investigated so far beyond the case of DP, where they appear
at one-loop order~\cite{oerding_dirperc} and they were measured
numerically in $d=1$~\cite{Gr-95}. 
These results imply that the average and variance of the first-crossing time
$t_{\rm cro}$ are both {\it finite} for DP in $d<4$ but {\it not} for the spin
models studied in Refs.~\cite{satya_clement_persist,oerding_persist}.

The evidence accumulated so far in two different models with $m_0 = 0$ and
$1$ show that the critical persistence
properties actually depend on $m_0$. It is therefore natural
to investigate how the corresponding crossover occurs when $m_0$ is
varied within the {\it same} model~\footnote{%
Note that in reaction-diffusion systems the state with $m_0=0$ (and no initial
fluctuations --- they would anyhow relax exponentially fast in time) correspond
to the inactive {\it absorbing} phase.} and which are the associated 
{\it universal} features.

\section{Universal scaling behavior of the persistence probability}
Here we focus primarily on the Ising model (on a $d$-dimensional hypercubic
lattice) with Glauber dynamics quenched to its critical
point and we study its persistence properties 
both analytically and numerically.  
The universal aspects of the relaxation of this model are
captured by the so-called Model A~\cite{hohenberg77} 
for the $n$-component fluctuating local order parameter $\varphi(x,t)$ (\eg,
the coarse-grained density of spins in the Ising model):
\begin{eqnarray}\label{def_Langevin}
\eta \partial_t \varphi(x,t) = - \frac{\delta
 \H[\varphi]}{\delta \varphi(x,t)} + \zeta(x,t)
\end{eqnarray}
where  $\zeta(x,t)$ is a Gaussian white noise
with $\langle \zeta(x,t) \rangle = 0$ and
$\langle \zeta(x,t) \zeta(x',t') \rangle = 2\eta T \delta(x-x')
\delta(t-t')$, $\eta$ is the friction
coefficient (set to $1$ in the following) and $T$ the temperature of the
thermal bath. In Eq.~(\ref{def_Langevin}), $\H$ is the $O(n)$-symmetric
Landau-Ginzburg functional which, for $n=1$, 
describes the universal aspects of the static properties of the Ising model:
\begin{equation}\label{def_O1}
\H[\varphi] = \int \rmd^d x \left[ \frac{1}{2}(\nabla \varphi)^2 +
 \frac{1}{2} r_0 \varphi^2 + \frac{g_0}{4!}
 (\varphi^2)^2 \right]  
\end{equation} 
where $r_0$ is a parameter which has to be 
tuned to a  critical value $r_{0,c}$ when approaching the critical temperature
$T_c$ (here $r_{0,c} = 0$) and $g_0 > 0$
is the bare coupling constant. 

At the initial time $t = 0$, the
system is in a random 
configuration with mean magnetization $[\varphi(x,0)]_i = M_0$ and 
short-range correlations $[\varphi(x,0) \varphi(x',0)]_i = \tau_0^{-1} \delta
(x-x')$ ($[\ldots]_i$ stands for the average over the distribution of the
initial configuration), where $\tau_0^{-1}$ is irrelevant in
determining the leading scaling properties~\cite{janssen_rg} 
(therefore we set $\tau_0^{-1}=0$). For $n=1$ (the case
$n>1$ is shortly discussed below), the
stochastic process we are interested in is 
\begin{equation}
\psi(x,t) = \varphi(x,t) - M(t)\,,\quad \mbox{where} 
\quad M(t) = \langle \varphi(x,t) \rangle 
\end{equation} 
is the average magnetization and $\langle\ldots\rangle$ stands for the average
over the possible realizations of the stochastic noise $\zeta$.
Defining $\tm(t)$ as 
\begin{equation}
\tm(t) = \frac{1}{L^{d}} \int\!\rmd^dx\, \psi(x,t) \label{def_mag_tilde}
\end{equation} 
we are interested in the probability $P_c(t)$ 
that $\tm$ does not change sign in the time interval $[0,t]$ following the
quench to $T_c$. 

We first present the result of the Gaussian approximation which is exact in
dimension $d>4$ and which is obtained by neglecting non-linear terms
in $\psi$ in the Langevin
Eq.~(\ref{def_Langevin}) expressed in terms of $\psi(x,t)$ and $m^2 \equiv g_0
M^2/2$~(see, \eg, Ref.~\cite{andrea_ordered_ising}):
\begin{equation}
\begin{split}
&\partial_t \psi(x,t) = [\nabla^2  - m^2(t)] \psi(x,t) + \zeta(x,t)
\\
&\mbox{where} \quad \partial_t m(t) + \frac{1}{3} m^3(t) = 0  \,.
\label{eq_gaussian}
\end{split}
\end{equation}
The equation of motion for $\tm$ [see Eq.~(\ref{def_mag_tilde})] 
readily follows:
$\partial_t \tm(t) + m^2(t) \tm(t) = \tilde \zeta(t)$, 
where $\tilde \zeta(t) = L^{-d}\int\!\rmd^d x\,\zeta(x,t)$ is still Gaussian and,
from Eq.~(\ref{eq_gaussian}), $m^2(t) = (2t/3 + m_0^{-2})^{-1}$
where $m(t=0) = m_0$. In the limit $t \ll m_0^{-2}$, one finds that
$\tm(t)$ executes a simple random walk and therefore
$P_c(t) \propto t^{-1/2}$~\cite{feller_book}. In the opposite 
limit $t \gg m_0^{-2}$, $\tm(t)$ satisfies the Langevin equation
$\partial_t \tm(t) + (3/2t) \tm(t) = \tilde \zeta(t)$
which turns into $\partial_\tau \hat
M(\tau)= \eta(\tau)$ when the variables $\hat M(t) = t^{3/2} \tm(t)$,
$\eta(t) = t^{-3/2}\tilde\zeta(t)/4$ and $\tau =
t^4$ are introduced~\cite{majumdar_critical}. Accordingly, 
$P_c(\tau) \propto \tau^{-1/2}$, \ie, $P_c(t) \propto t^{-2}$, which decays
to zero more rapidly than for $t\ll m_0^{-2}$ and clearly shows the existence
of two different regimes. The full crossover function can be obtained 
by noticing that $\tm(t)$ is a Gaussian process and therefore
it is completely
determined by its two-time correlation function
$C_\tm(t,t')=\langle\tm(t)\tm(t')\rangle$ which can be easily
calculated~\cite{andrea_ordered_ising}. Following
Ref.~\cite{majumdar_critical}, 
one introduces the normalized process 
$X(t) = \tm(t)/\langle \tm^2(t) \rangle^{1/2}$, the two-time correlation
function of which is given by ($t>t'$)~\cite{andrea_ordered_ising}:
\begin{equation}\label{expr_gauss_complete}
\begin{split}
&\langle X(t) X(t') \rangle =
\frac{C_{\tm}(t,t')}{\sqrt{C_{\tm}(t,t)C_{\tm}(t',t')}}= \frac{L(\tilde
  t')}{L(\tilde t)} \\
&\mbox{where}\quad L(\tilde t) = \sqrt{(\tilde t+1)^4 - 1}
\end{split}  
\end{equation}
and the dimensionless time variables are given by $\tilde t =
t/\tau_m$ with $\tau_m = 3/(2 m_0^2)$. In terms of the logarithmic time
$T = \ln{L(t)}$, $X(T)$ is a stationary Gaussian process with purely
exponential correlations $\langle X(T) X(T')\rangle = \exp{(-|T-T'|)}$, for
which the persistence probability is known
exactly~\cite{slepian,satya_review}. 
In the limit $t,\tau_m \gg t_{\rm micr}$ (where $t_{\rm micr}$ is some non-universal
microscopic time scale)  one finds
\begin{eqnarray}\label{crossover_gaussian}
P_c(t) \propto \frac{L(\tilde t_{\rm micr})}{L(\tilde t)} &\simeq&
\frac{(\tau_m/t_{\rm micr})^{-1/2}}{\sqrt{(1+t/\tau_m)^4 - 1}} \nonumber \\
&\sim&
\begin{cases}
t^{-\frac{1}{2}} \,,& t \ll \tau_m \\
t^{-2}\,, & t \gg \tau_m
\end{cases} 
\end{eqnarray}   
in agreement with the results obtained by mapping the process on a random
walk. 
The mean first-crossing time $t_{\rm cro}$ resulting from 
$P_c(t)$ is {\it finite} for $m_0\neq 0$,
whereas its variance {\it is not}.
Although the Gaussian approximation is exact only for $d > 4$, it clearly
displays for finite $m_0$ (\ie, finite $\tau_m$) 
an interesting crossover~(\ref{crossover_gaussian}) --- which is
expected to occur for generic $d$ above the lower critical dimension of
the model ---
between two successive regimes in which $P_c(t)$ decays algebraically with two
different exponents, $\theta_0$ and $\theta_\infty$, respectively.
The first regime ($t \ll \tau_m$) has a temporal extent $\sim\tau_m$ which increases upon decreasing $m_0$ and indeed, in the limit $m_0\rightarrow
0$ considered in Refs.~\cite{majumdar_critical,oerding_persist}, 
it is the only one accessible. 
The crossover to the second regime takes place for $t\simeq \tau_m$, with
$\tau_m$ decreasing upon increasing $m_0\neq 0$ and indeed $\tau_m$ tends to
zero in the case $m_0\rightarrow\infty$ considered in
Ref.~\cite{oerding_dirperc} for the DP.
%
%
%To get some insight on this crossover in lower dimension, $d < 4$, 
%it is natural to study the simple case of $d=1$ Ising model
%with Glauber dynamics, at $T=0$ and a non zero initial magnetization
%$m_0$. However, in that case one can show easily from Ref. \cite{amar_family}
%that $P_c(t) \propto t^{-1/4}$, independently of $m_0$. Notice that this is
%consistent with the monotonic behavior - in particular no initial slip
%behavior-of the magnetization $\langle M_0(t) 
%\rangle = m_0$, $\forall t$. 

To go beyond the Gaussian approximation, we   
take advantage of recent results about the aging behavior of systems
described by Eqs.~(\ref{def_Langevin}) and (\ref{def_O1}) following 
a quench from a state with $m_0\neq 0$ to the critical 
point~\cite{andrea_ordered_ising}.
As mentioned in the introduction, $\tilde
M(t)$ (and therefore $X(t) = \tm(t)/\langle \tm^2(t)\rangle^{1/2}$) 
is a Gaussian process in the
thermodynamic limit. 
Taking into account the scaling form of $C_\tm(t,t')$ discussed in 
Ref.~\cite{andrea_ordered_ising}, one expects for $t> t'$:
\begin{eqnarray}\label{x_gen}
\langle X(t) X(t') \rangle = \left(t/t'\right)^{-\mu_0}
F_X(t'/t,t/\tau_m) 
\end{eqnarray}
where $\tau_m \equiv (B_mm_0)^{-1/\kappa}$,
$B_m$ is a non-universal constant which can be fixed according to some
normalization condition (see Ref.~\cite{andrea_ordered_ising} for details), 
$F_X$ is a {\it universal} scaling function (with $F_X(1,y)=1$) and $\mu_0 =
-\theta' + (1-\eta/2)/z$.
For $t/\tau_m \ll 1$ one recovers the
result of Refs.~\cite{majumdar_critical,oerding_persist}. However, a
non-vanishing mean $m_0$ 
of the initial order parameter affects the behavior of
the temporal correlations as soon as $t \sim \tau_m $ and in
particular for $t, t' \gg \tau_m$. The limiting behaviors of
Eq.~(\ref{x_gen}) can be derived from those of $C_\tm(t,t')$ discussed in
Ref.~\cite{andrea_ordered_ising}:
\begin{eqnarray}\label{x_gen_crossover}
\langle X(t) X(t') \rangle \sim 
\begin{cases}
(t/t')^{-\mu_0} f_0(t'/t)\,, &t'< t \ll \tau_m \\
(t/t')^{-\mu_\infty} f_\infty(t'/t)\,, &t>t' \gg \tau_m
\end{cases}
\end{eqnarray}
where $f_0(x) = F_X(x,0)$, $f_\infty(x) = x^{\mu_0-\mu_\infty} F_X(x,y\rightarrow\infty)$ with $f_{0,\infty}(0)$ finite [$f_{0,\infty}(1)=1$]
and, for $d<4$, $\mu_\infty = 1+d/(2z)$\footnote{% 
Note that this relation is the same as the one  derived in
Ref.~\cite{oerding_persist} for the DP universality class.}. % 
Note that if $f_{0,\infty}(x)$ is {\it constant} 
for $x\in[0,1]$, $\langle X(t) X(t') \rangle$ can be cast
asymptotically in the
form~(\ref{expr_gauss_complete}) with a suitable choice of $L(t)$ and
therefore the process $X(t)$ can be mapped into a Markovian one
with $P_c(t_{\rm micr}\ll t \ll \tau_m) \sim t^{-\mu_0}$ and 
$P_c(t\gg\tau_m) \sim t^{-\mu_\infty}$.
In the scaling limit, the relevant time scales for the problem of persistence
are the time $t$ elapsed since the quench and the (mesoscopic) time scale
$\tau_m$ set by the value of the initial magnetization. Accordingly, one
expects (also on the basis of the Gaussian approximation) a scaling behavior 
\begin{eqnarray}\label{eq_crossover}
P_c(t) = A_P t^{-\theta_0} {\cal P}(t/\tau_m)\,, \quad
{\cal P}(x\gg 1) \sim x^{-\theta_\infty+\theta_0} 
\end{eqnarray}   
where $A_P$ is a non-universal constant such that ${\cal P}(0) = 1$, 
${\cal P}(x)$ is a
universal scaling function, $\tau_m$ is fixed by the behavior of
the magnetization as in Ref.~\cite{andrea_ordered_ising}, 
and $\theta_\infty$ is a yet undetermined exponent. 
Equation~(\ref{eq_crossover}) implies $\langle t_{\rm cro}\rangle =
\int_0^\infty \!\rmd t P_c(t) \sim
\tau_m^{1-\theta_0}$ and therefore
the first-crossing time $t_{\rm cro}$ has a
finite average for $\theta_0<1<\theta_\infty$
and a finite second moment 
$\langle t^2_{\rm cro}\rangle = 2 \int_0^\infty \!\rmd t\, t P_c(t) 
\sim \tau_m^{2-\theta_0}$
for $\theta_0<2<\theta_\infty$.  
Note that $\theta_{0,\infty} =
\mu_{0,\infty}$ for a Markovian process.   
The corrections to $\theta_0$ due to non-Markovian effects
(reflected in the fact that $f_{0,\infty}(x)$ in Eq.~(\ref{x_gen_crossover}) 
are not constant) were obtained at two-loop order 
in Ref.~\cite{oerding_persist}. 
For $\theta_\infty$ these corrections have not been calculated. 
They can be determined by focusing on the case $m_0 \to\infty$ of
Eq.~(\ref{x_gen_crossover}), \ie, on $f_\infty(x)$, which was calculated in
Ref.~\cite{andrea_ordered_ising} up to one loop in the $\epsilon$-expansion
($\epsilon = 4 - d$). Taking advantage of
the results therein one 
finds, in logarithmic time $T = \ln t$ ($T > T'$),
\begin{eqnarray}\label{correl_A}
&&\langle X(T) X(T') \rangle = \rme^{-\mu_\infty(T-T')}
{\cal A}(\rme^{-(T-T')}) ,\\
&&\mbox{where} \quad {\cal A}(x) = 1 + \epsilon{\cal A}_1(x) + {\cal O}(\epsilon^2)\,,  \nonumber
\end{eqnarray}       
${\cal A}_1(x) = 2[f_C(x) - f_C(1)]/3$ and $f_C(x)$ is given in
Eq.~(B.31) of Ref.~\cite{andrea_ordered_ising}. The universal function 
${\cal A}_1(x)$ carries the signature of the non-Markovian corrections to
$\theta_\infty$ which, in turn, can be calculated by using the perturbation theory of
Ref.~\cite{oerding_persist} (see Eq.~(14) therein):
\begin{eqnarray}\label{theta_oneloop}
{\mathcal R} &\equiv& \frac{\theta_\infty}{\mu_\infty} = 1 - \epsilon
\frac{2\mu_\infty}{\pi} \int_0^1 \!\rmd x \frac{x^{\mu_\infty-1} {\cal
    A}_1(x)}{(1 -  x^{2\mu_\infty})^{3/2}} 
+ {\cal O}(\epsilon^2) \nonumber \\
&=& 1 + \epsilon \, 0.0273.. + {\cal O}(\epsilon^2)
\end{eqnarray}
where $\mu_\infty = 1 + d/(2 z) = 2 (1-\epsilon/8) + {\cal O}(\epsilon^2)$ 
and the integral has
been computed numerically. At variance with %the case of 
$\theta_0$ for Model A,
non-Markovian corrections appear already at one-loop order, as in the
DP and Model C~\cite{oerding_dirperc,oerding_persist}.  
Compared to its Markovian and Gaussian approximations,
the resulting exponent 
$\theta_\infty = 2 -  \epsilon\, 0.195.. + {\cal O}(\epsilon^2)$ is
respectively increased and decreased. 
Accordingly, $t_{\rm cro}$ has a finite average $\langle t_{\rm cro} \rangle
\sim m_0^{-(1-\theta_0)/\kappa}$ which diverges with a universal
exponent upon reducing $m_0$. In the opposite limit
$m_0\rightarrow\infty$, instead, $\langle t_{\rm cro} \rangle$ 
is finite and determined by $t_{\rm micr}$ 
(as for the DP~\cite{oerding_dirperc}).

It is also instructive, 
though less relevant for the description of %
systems other than
lattice models, to calculate the persistence
probability of a $n$-component vector order parameter 
$\varphi = (\varphi_1,
\varphi_2,..,\varphi_n)$ with $n>1$ and Model A dynamics (see
Eqs.~(\ref{def_Langevin}) and~(\ref{def_O1})).
When the initial magnetization $m_0$ is finite, 
fluctuations which are parallel
$\psi_{\sigma}(x,t)$
and transverse $\psi_{\pi}(x,t)$ (with $O(n-1)$ symmetry) to the average order
parameter $m(t) = \langle \varphi(x,t)\rangle$
are expected to have 
distinct persistence 
probabilities, $P^{\sigma}_c(t)$ and $P^{\pi}_c(t)$, respectively. 
Taking advantage of the
results of Ref.~\cite{andrea_ordered_on} one finds
that both of them exhibit a crossover as in
Eq.~(\ref{eq_crossover}) with $n$-dependent exponents 
$\theta^{\sigma}_0(n) = \theta^{\pi}_0(n)$ (for $t\ll\tau_m$ the 
$O(n)$-symmetry is
restored and the results of 
Ref.~\cite{oerding_persist} apply) 
and $\theta^{\sigma,\pi}_\infty(n)$ which in the
Markovian approximation read $\mu_\infty^\sigma = 1 + d/(2 z)$ and
$\mu_\infty^\pi = 1 - d/(2 z) + 2\beta/(\nu z)$. Non-Markovian corrections
contribute already at one loop, yielding ${\mathcal R}^\sigma = 1 +
\epsilon(0.115.. + 
n\, 0.131..)/(8+n)+ {\cal O}(\epsilon^2)$ and  
${\mathcal R}^\pi = 1 + \epsilon\,0.055../(8+n)
+ {\cal O}(\epsilon^2)$ (an expansion of ${\mathcal R}^\pi$ around $d=2$ can
be obtained from the analysis of Ref.~\cite{fedo}).
Interestingly enough, by taking
advantage of the results of Ref.~\cite{andrea_ordered_on},
$P^\pi_c(t)$ can be computed exactly for
$n\rightarrow\infty$ and $2<d<4$.
In the short-time regime $t\ll\tau_m$,
one recovers 
$\theta^\pi_0(n \to \infty) = (d-2)/4$, \ie, the same expression as for the
case $m_0=0$~\cite{majumdar_critical} whereas,
for $t \gg \tau_m$, one finds $\theta_\infty^\pi (n \to \infty) = d/4$. In
both cases there are no non-Markovian corrections and indeed the correlation
function of the normalized
process $X^\pi(t)$ associated to $\tm_\pi(t) = L^{-d}\int\!\rmd^d
x\,\psi_\pi(x,t)$ takes the form~(\ref{expr_gauss_complete}) with $L(\tilde t)
= \tilde t^{(d-2)/4}[\tilde t + d/(d-2)]^{1/2}$ and $\tau_m = 3(d-2)/(4
m_0^2)$, allowing also the calculation of the full cross-over function.
In passing we mention that a similar crossover occurs also in the spherical
model with purely dissipative dynamics: %
By taking advantage of the results of Ref.~\cite{as-06} we find 
$\theta_\infty = \mu_\infty = d/4 + 1$ (in agreement with the hyperscaling
relation) which differs indeed from $\theta_0 = \mu_0 =
(d-2)/4$~\cite{majumdar_critical}.

\section{Monte Carlo simulations of the two-dimensional Ising model}
To test the expected scaling behavior (see Eq.~(\ref{eq_crossover})),
the corresponding crossover as well as our theoretical estimate of
$\theta_\infty$, we performed Monte Carlo simulations of the
two-dimensional Ising model (with spins $s_i = \pm 1$) on a $L\times
L$ lattice with periodic boundary conditions and Glauber dynamics.
The system is prepared in a random configuration with 
$N_{+}$ up and $N_{-}$ down spins, where $N_{\pm} = L^2 (1 \pm
m_0)/2$. 
At each time step, one site is randomly
chosen and the move $s_i \mapsto - s_i$ is accepted or rejected according to
Metropolis rates. One time unit
corresponds to $L^2$ such time steps. 
The computation of the persistence probability $P_{c}(t)$ requires the
knowledge of the %
magnetization 
$M(t) = L^{-2} \langle \sum_i s_i\rangle$, 
which we averaged over 2000 thermal 
realizations. After the quench to $T_c$, 
each system evolves until $\tm(t) = L^{-2}  \sum_i s_i - M(t)$ 
first crosses zero. ${P}_c(t)$
is then measured as the fraction 
of surviving systems at each time $t$, 
computed by averaging over
$10^5$ samples. 
\begin{figure}
\includegraphics[scale=0.3]{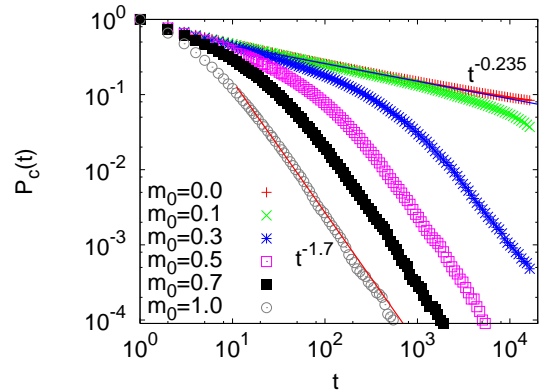}
\caption{Persistence probability $P_c(t)$ of the two-dimensional
critical 
Ising model with Glauber dynamics, for different initial magnetization 
$m_0$. All the curves decay initially with an exponent 
$\theta_0\simeq 0.235$ and then, after the crossover, 
with a different exponent $\theta_\infty\simeq 1.7$.}
\label{fig1}       
\end{figure}
In Fig.~\ref{fig1} we plot $P_c(t)$ for $L=256$ 
and 
different values of $m_0$ ranging from $0$ to $1$. The curves
with $m_0 > 0.1$ display a clear crossover between two different algebraic
decays, as anticipated by our theoretical analysis. 
We have carefully checked --- by 
computing $P_c(t)$ for $L$ ranging between $64$ and $512$ --- that
the displayed crossover is not affected by finite-size effects.  
For small enough $t$, $P_c(t)$ decays with an exponent
$\theta_0^{\text{(MC)}}=0.235(5)$ (fully compatible with the available
estimate~\cite{Schulke97}),
whereas for larger times the power-law decay is 
faster, with 
$\theta_\infty > \theta_0$ and  
\begin{eqnarray}\label{theta_numerics}
\theta_\infty^{(\text{MC})} = 1.7(1)\,.
\end{eqnarray} 
An analytical estimate of this exponent 
can be obtained by setting
$\epsilon=2$ in Eq.~(\ref{theta_oneloop}), 
yielding $\theta_\infty^{\rm (1loop)} \simeq 1.61$
which is actually in rather good agreement
with $\theta_\infty^{(\text{MC})}$.
As expected from the scaling form~(\ref{eq_crossover}),
the time $\tau_m$ at which the crossover occurs %in the curves reported 
in
Fig.~\ref{fig1} increases as the initial magnetization $m_0$ decreases and
eventually no crossover is observed for
$m_0=0$~\cite{majumdar_critical}. 
\begin{figure}
\includegraphics[scale=0.3]{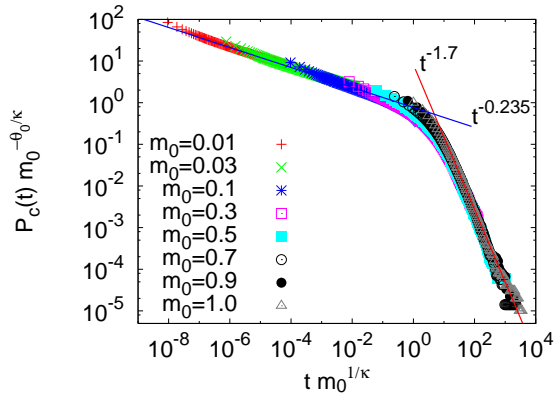}
\caption{Plot of $P_c(t) m_0^{-\theta_0/\kappa}$ as a function of the
 rescaled time $t m_0^{1/\kappa}$ with $\theta_0=0.235$ and $\kappa =
0.249$, including the data 
 presented in Fig.~\ref{fig1}. According to Eq.~(\ref{eq_crossover})
the exponent of the ultimate power-law decay is $\theta_\infty\simeq
1.7$. 
} \label{fig2}
\end{figure}
In Fig.~\ref{fig2} we plot $P_c(t) m_0^{-\theta_0/\kappa}$
($\propto P_c(t)\tau_m^{\theta_0}$)  as a function of the rescaled time $t
\, m_0^{1/\kappa}$ ($\propto t/\tau_m$) in order to highlight the scaling
behavior~(\ref{eq_crossover}).
The exponent $\kappa$ is related to 
$\beta=1/8$, $\nu=1$, $\theta' = 0.191(3)$~\cite{Gr-95} and
$z=2.1667(5)$~\cite{nightingale_z} via the
scaling relation $\kappa = \theta' + \beta/(\nu z)$~\cite{janssen_rg}, which
yields $\kappa \simeq 0.249$. For 
$\theta_0$ we use
the estimate $\theta_0^{\text{(MC)}}$ obtained from
Fig.~\ref{fig1}. The quite good data collapse in Fig.~\ref{fig2}
is remarkably obtained without 
fitting parameters and therefore
it provides a convincing numerical 
evidence of the scaling behavior~(\ref{eq_crossover}).

\section{Conclusions}
We have studied the global persistence probability $P_c(t)$ for critical
system which are initially prepared in a state with
short-range correlations and non-vanishing average $m_0$ of the order
parameter. For different models, we have shown that $P_c(t)$ exhibits a
crossover (apparently overlooked in previous studies) between two distinct
power-law behaviors which is described by a universal scaling
function ${\cal P}$ (see Eq.~(\ref{eq_crossover})). 
Our Monte Carlo simulations of the two-dimensional 
Ising model with Glauber dynamics clearly display this crossover. 
We have calculated non-Markovian corrections to the (universal) 
persistence exponent
$\theta_\infty$ which characterizes the ultimate long-time behavior of 
$P_c(t)$ for $m_0 > 0$ and it turns out to be in good agreement with our
numerical estimates. 
It would be interesting to extend beyond the Gaussian approximation 
the calculation of the scaling function ${\cal P}$.

In this paper we have studied the crossover in the persistence of the 
{\it global} magnetization $M$ of a $d$-dimensional system at criticality. In
addition one might have considered the magnetization $\tilde M_{\rm sub}$ 
of a $d'$-dimensional submanifold of the original system 
($0\le d'\le d$)~\cite{manifold} 
or the magnetization $M_V(t)$ of a subsystem of finite 
volume $V = \ell^d$~\cite{sire}.
It turns out that, depending on $d'$, the long-time decay of the 
persistence probability of 
$\tilde M_{\rm sub}$ is exponential,
stretched exponential, or algebraic~\cite{manifold}, whereas the persistence
probability of $\tilde M_V$ is algebraic for $\xi(t) \ll \ell$ and
exponential for $\xi(t) \gg \ell$.
 
The dependence on $d'$ and $\ell$ of the 
dynamic crossover highlighted in this paper and the interplay among the
various crossovers surely deserve further investigations.

\end{document}